%% file: main.tex
\documentclass[conference]{IEEEtran}
\IEEEoverridecommandlockouts
\usepackage{cite}
\usepackage{amsmath,amssymb,amsfonts}
\usepackage{algorithmic}
\usepackage{graphicx}
\usepackage{textcomp}
\usepackage{xcolor}
\usepackage{comment}
\usepackage{hyperref}
\usepackage{threeparttable}
\usepackage[font=footnotesize]{caption}
\usepackage{afterpage}
\usepackage{float}
\usepackage{subcaption} 

\usepackage{balance}

\def\BibTeX{{\rm B\kern-.05em{\sc i\kern-.025em b}\kern-.08em
    T\kern-.1667em\lower.7ex\hbox{E}\kern-.125emX}}
\begin{document}

\title{Real-Time Detection of Robot Failures Using Gaze Dynamics in Collaborative Tasks
}

\author{\IEEEauthorblockN{Ramtin Tabatabaei}
\IEEEauthorblockA{\textit{The University of Melbourne} \\
Melbourne, Australia \\
stabatabaeim@student.unimelb.edu.au}
\and

\IEEEauthorblockN{Vassilis Kostakos}
\IEEEauthorblockA{\textit{The University of Melbourne} \\
Melbourne, Australia \\
vassilis.kostakos@unimelb.edu.au}
\and

\IEEEauthorblockN{Wafa Johal}
\IEEEauthorblockA{\textit{The University of Melbourne} \\
Melbourne, Australia \\
wafa.johal@unimelb.edu.au}}


\maketitle

\begin{abstract}

Detecting robot failures during collaborative tasks is crucial for maintaining trust in human-robot interactions. This study investigates user gaze behaviour as an indicator of robot failures, utilising machine learning models to distinguish between non-failure and two types of failures: executional and decisional. Eye-tracking data were collected from 26 participants collaborating with a robot on Tangram puzzle-solving tasks. Gaze metrics, such as average gaze shift rates and the probability of gazing at specific areas of interest, were used to train machine learning classifiers, including Random Forest, AdaBoost, XGBoost, SVM, and CatBoost.
The results show that Random Forest achieved 90\% accuracy for detecting executional failures and 80\% for decisional failures using the first 5 seconds of failure data. Real-time failure detection was evaluated by segmenting gaze data into intervals of 3, 5, and 10 seconds. These findings highlight the potential of gaze dynamics for real-time error detection in human-robot collaboration.


\end{abstract}


\input{sections/01_Introduction}
\input{sections/02_RelatedWorks}

\input{sections/03_Methodology}

\input{sections/04_Results}

\input{sections/05_Discussion}

\section*{Acknowledgement}
This research is partially supported by the Australian Research Council Discovery Early Career Research Award (Grant No. DE210100858)

\newpage

\bibliographystyle{ieeetr}
\balance
\bibliography{references,hri2025}

\end{document}

%% file: sections/01_Introduction.tex
\section{Introduction}

The potential for robots to assist people in various domains is becoming increasingly evident \cite{sauppe_social_2015,  babel_step_2022, chatterjee_usage_2024}. They can collaborate with humans as teammates to perform joint activities \cite{mingyue_ma_human-robot_2018}. To ensure successful collaboration, it is crucial for robots to exhibit effective behaviour and communication, as this helps maintain alignment and fosters trust \cite{ desai_effects_2012}. However, as robots become more integrated into daily life, their inevitable errors—caused by real-world uncertainties—pose risks to task success, user safety, and trust \cite{schaefer_meta-analysis_2016, salem_would_2015, sebo_i_2019}. Trust in human-robot collaboration fluctuates, dropping after failures but recovering if the robot quickly detects and corrects its mistakes \cite{lemasurier_reactive_2024, wachowiak_when_2024, kraus_sorry_2023}.  To recover effectively from errors, robots should not only detect their failures but also identify the specific type of failure (e.g., motion execution versus task planning). Different types of failures require specific recovery approaches \cite{wachowiak_when_2024}, making accurate failure identification a key capability for robots in collaborative settings. One promising strategy for enabling robots to detect their own failures is by modelling user reactions during the moment of failure. This involves analysing signals such as social and non-verbal cues, with eye gaze emerging as a particularly valuable indicator \cite{wachowiak_analysing_2022}. Eye gaze conveys information about attention \cite{velichkovsky_social_2021, fang_dual_2021}, and emotional states \cite{velichkovsky_social_2021, huang_using_2015}. By leveraging machine learning algorithms to model user gaze behaviour, robots can monitor gaze patterns to detect failures in real-time, improving their ability to respond effectively and maintain trust.

This study explores the development of machine learning classifiers to detect robot failures using user gaze patterns during collaborative tasks. It focuses on two research questions: (RQ1) how the performance of these models varies based on the time elapsed after a robot failure, and (RQ2) how the performance of these models varies when applied to real-time failure detection.

To address these questions, we used data collected on a total of 26 participants engaged in four sessions of Tangram puzzle-solving, during which the robot was intentionally programmed to fail once per puzzle \cite{ramtin_hri_2025}. The results of the machine learning classifiers show that the models perform well in detecting failures. When implemented in real-time, they can detect most failures effectively.

%% file: sections/02_RelatedWorks.tex
\section{RELATED WORKS}

Research has shown that users display common instinctive social signals during robot errors, distinguishing these situations from error-free scenarios. These signals include gaze behaviour \cite{kontogiorgos_embodiment_2020, peacock_gaze_2022, kontogiorgos_systematic_2021, ramtin_hri_2025}, facial expressions \cite{mirnig_impact_2015, stiber_modeling_2022, kontogiorgos_systematic_2021, wachowiak_time_2024}, verbalisation \cite{kontogiorgos_embodiment_2020, mirnig_impact_2015, kontogiorgos_systematic_2021}, and body movements \cite{mirnig_impact_2015, trung_head_2017, wachowiak_time_2024}. For example,  Peacock et al. \cite{peacock_gaze_2022} observed that gaze initially increases in motion during failures and then stabilises as users address the issue. Stiber et al. \cite{stiber_using_2023} identified heightened activity in facial muscles, such as smiling and brow lowering, during robot errors. Similarly, Kontogiorgos et al. \cite{kontogiorgos_systematic_2021, kontogiorgos_embodiment_2020} reported increased spoken words, longer utterances, and more gaze shifts toward the robot, reflecting greater user engagement during failures.

Several studies have explored machine-learning approaches to detect failures in human-robot interactions using various behavioural and physiological cues. For example, Peacock et al. \cite{peacock_gaze_2022} trained logistic regression models on gaze dynamics to detect failures, achieving accurate detection a few seconds after the errors occurred. Similarly, Kontogiorgos et al. \cite{kontogiorgos_systematic_2021, kontogiorgos_behavioural_2020} developed machine learning models, including XGBoost and Random Forest, that utilised multimodal behaviours—such as linguistic, facial, and acoustic features—to achieve high accuracy in distinguishing failure scenarios from non-failure scenarios in a verbal guidance scenario. Separately, Stiber et al. \cite{stiber_modeling_2022} trained Multi-Layer Perceptron (MLP) models using action units (AUs) derived from facial reactions, showing that AUs are effective if users provide timely and observable responses to robot errors. Since not all users exhibit clear facial or verbal reactions to failures, this highlights a gap in the literature. This research aims to address this gap by designing classifier models based on user gaze during a collaborative task, evaluating their performance relative to the time elapsed after a robot failure, and assessing their effectiveness in real-time settings.

%% file: sections/03_Methodology.tex
\section{Methodology}

\subsection{Tasks Description}

The experiment consisted of four collaborative tasks where a participant and a robot worked together to solve Tangram puzzles. In each task, participants were tasked with creating a unique shape using Tangram pieces. To avoid any influence of task difficulty on participant perception or behaviour, puzzles of comparable difficulty were carefully selected (See \cite{ramtin_hri_2025} for more details on the experimental setup).

Each Tangram puzzle consisted of seven pieces. The robot was responsible for placing four pieces (two small triangles, a square, and a parallelogram), while the participant placed the remaining three pieces (a medium triangle and two large triangles). The Tangram pieces were 3D-printed. The silhouettes of the puzzles were printed in black on A2-sized white paper, which was fixed to the table. Both the participant and the robot placed their pieces into the Tangram figure. Participants were instructed to move their pieces only after the robot had completed its action, with the robot always beginning by placing the first piece.

The robot's pieces were positioned near its workspace, adjacent to the paper.
For each puzzle, the robot determined the placement and orientation of its pieces before initiating movement. The Tiago robot, programmed using ROS1, utilized the `tf` library to map the pose of each piece to the coordinate frame of its robotic arm. It then employed inverse kinematics to precisely position its arm for accurate placement of each piece.


\subsection{Robot Failures}

We designed the robot to deliberately fail during each task, with failures categorized into two types: Executional Failure (EF) and Decisional Failure (DF), representing common technical issues robots may face in interactions.

\begin{itemize}
    \item \textbf{EF:} The robot pauses for 15 seconds after grasping an object, holding it in its end effector during the pause. After the delay, it resumes the task and completes the pick-and-place action.
    \item \textbf{DF:} The robot picks up an object but mistakenly moves to the location intended for a different object. It places the object incorrectly, pauses for 5 seconds, and then corrects the mistake by lifting the object and placing it in the correct location.
\end{itemize}

During both failure and non-failure events, the robot followed the same pick-and-place procedure, with failures differing only in task duration: EFs added a 15-second pause, while DFs added 16.5 seconds due to incorrect placement and correction.

To avoid timing biases, the robot's malfunctions were programmed to occur at different points in each task. Specifically, failures could occur either at the beginning of the collaboration, while the robot was placing the first piece, or toward the end of the interaction, while the robot was placing the third piece.

\subsection{Experiment}

A total of 26 participants (16 females, 9 males, and 1 non-binary; aged 18–34) were recruited from a university platform. They provided informed consent, received gift vouchers as compensation, and were debriefed after the experiment.

Participants were guided by an experimenter who explained the tasks and intervened as needed to ensure safety or trigger the robot’s responses. Participants wore eye-tracking glasses to record their gaze data.

Each participant completed four puzzles, with each puzzle lasting approximately 3 minutes, followed by a short break between puzzles. The study was conducted in a controlled laboratory setting with participants who had no prior experience with robotics. In each puzzle, the robot was responsible for placing four pieces, correctly placing three and making a failure with one. This failure was pre-programmed to vary by type (Executional or Decisional) and timing (Early or Late). These combinations were counterbalanced using a four-condition Latin-Square design, ensuring balanced exposure across conditions and minimizing timing effects.

\begin{table}[h!]
\centering
\resizebox{0.48\textwidth}{!}{
\begin{tabular}{ c|c c c c c}

\textbf{Participant ID}  & \textbf{Puzzle 1} & \textbf{Puzzle 2} & \textbf{Puzzle 3} & \textbf{Puzzle 4} \\ \hline
1  & EF (Early) & EF (Late) & DF (Late) & DF (Early) \\ 
2  & EF (Late) & DF (Early) & EF (Early) & DF (Late)\\ 
3  & DF (Early) & DF (Late) & EF (Late) & EF (Early)\\ 
4  & DF (Late) & EF (Early) & DF (Early) & EF (Late)\\
5  & EF (Early) & EF (Late) & DF (Late) & DF (Early)\\
...  &  ... &  ... & ... & ...\\ 
14  & EF (Early) & EF (Late) & DF (Late) & DF (Early) \\ 
...  &  ... &  ... & ... & ... \\
\end{tabular}

}
\caption{Order of failure type with their timing across puzzles}
\label{table:participant_data}
\end{table}

\subsection{Measures}\label{Measures}




For each puzzle and piece, the robot's actions (e.g., moving or picking up objects), failure occurrences, and failure types were recorded, along with participant gaze data collected using Neon Eye Tracking Glasses. For more details on the methodology, refer to Paper \cite{ramtin_hri_2025}.

Using the gaze data, we calculated several metrics, including the average rate of gaze shifts towards all AOIs, the average rate of gaze shifts towards the robot's body, the average duration of gaze directed at the end effector, the probability of gazing at each AOI, transition entropy, and stationary entropy. For successful pickups and placements, metrics were calculated from when the robot began picking up the object until it placed the piece. For EFs, metrics covered the 15-second failure period, while for DFs, they spanned from the robot's movement towards the incorrect location to completing the placement and pausing for 5 seconds.

Using gaze behaviour metrics, machine learning models were trained to classify each type of failure against a no-failure condition in a binary manner. Each participant contributed 12 data points for non-failure conditions, 2 data points for EFs, and 2 data points for DFs. Five classifiers were employed: Random Forest, configured with 100 decision trees; AdaBoost, with 100 boosting iterations; XGBoost, performing 100 boosting rounds with a learning rate of 0.01; Support Vector Machine (SVM), using a linear kernel; and CatBoost, configured with 100 iterations, a learning rate of 0.1, and a tree depth of 6. Classifiers were implemented using Scikit-learn, XGBoost, and CatBoost libraries.

To address data imbalance and prevent bias, we applied SMOTE normalization with k-neighbors set to 2. The trained models were evaluated using two approaches. First, we assessed their performance in distinguishing failure events from non-failure events based on the first \textit{n} seconds of a failure. Second, to evaluate real-time failure detection, we analysed the models' performance in identifying failure types by segmenting the eye-tracking data into intervals of 3, 5, and 10 seconds, using a sliding window of 1 second. For both evaluation methods, we employed leave-one-out cross-validation, where models were trained on data from 25 participants and tested on the data from the remaining participant.

The primary goal of the classifiers was to achieve high accuracy while minimising false negatives, as failing to detect a failure event is critical in this context. Given this aim, we focus on reporting only the accuracy and recall metrics.

%% file: sections/04_Results.tex
\section{RESULTS}

\subsection{Evaluating Classification Performance with Varying Failure Times}
  

\begin{figure}[H]
  \centering
  \begin{subfigure}[t]{0.48\linewidth}
    \centering
    \includegraphics[width=\linewidth]{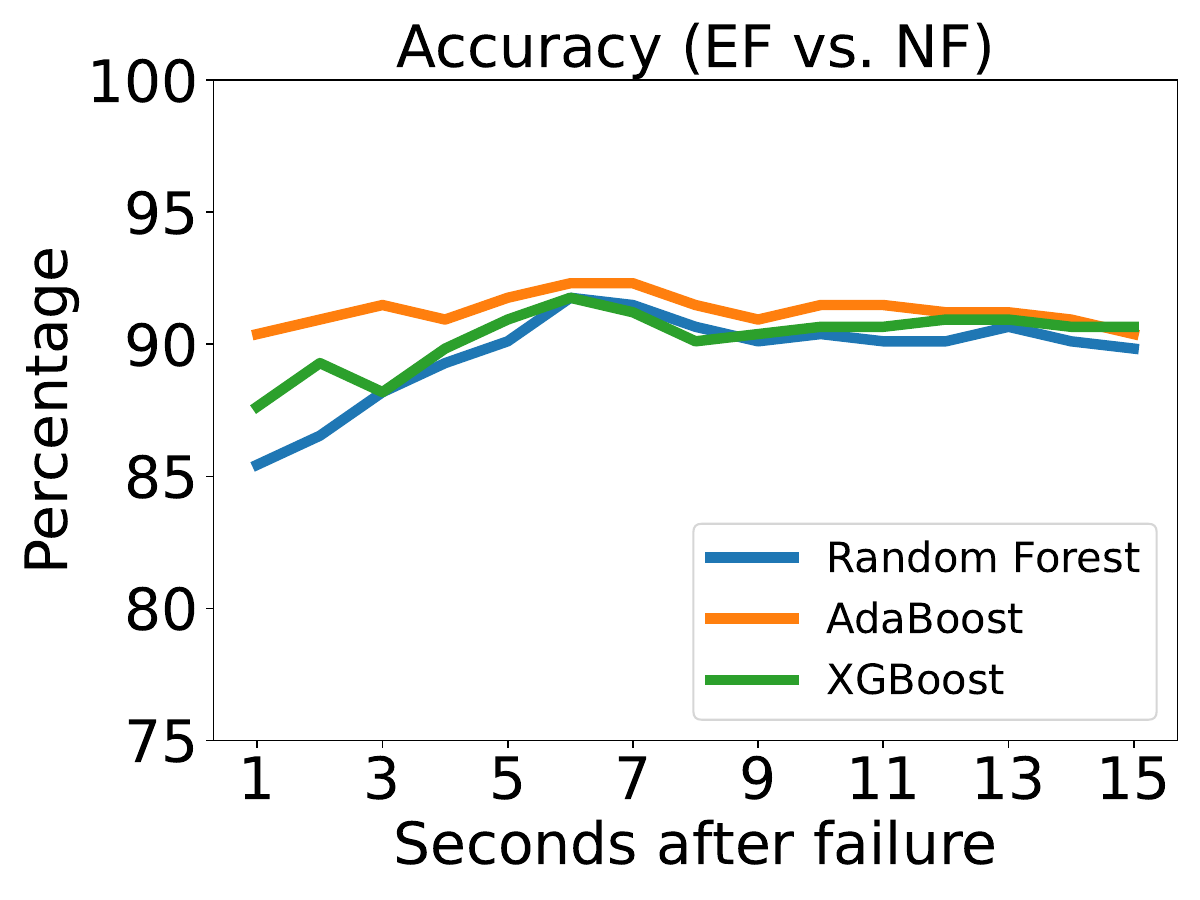}

    \label{GazeShiftsAllAoIs}
    \caption{ }
  \end{subfigure}
  \hfill
  \begin{subfigure}[t]{0.48\linewidth}
    \centering
    \includegraphics[width=\linewidth]{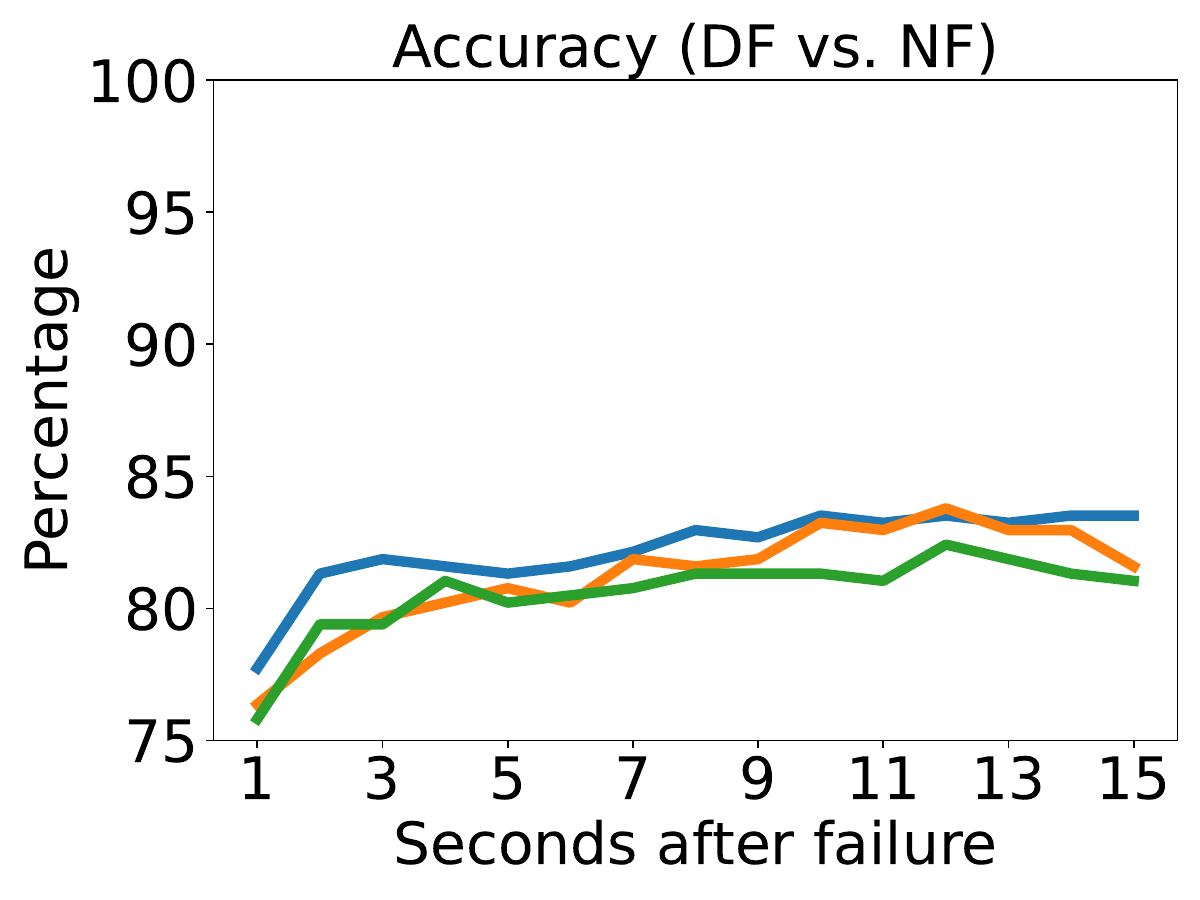}

    \label{GazeShiftsRobotBody}
    \caption{ }
  \end{subfigure}
  \caption{Classifier accuracy assessed using the first \textit{n} seconds of the failure period: a) distinguishing between non-failure and executional failure, and b) distinguishing between non-failure and decisional failure.}
  \label{fig2}
\end{figure}

To address the first research question, we evaluated the trained models (Random Forest, AdaBoost, and XGBoost), which were trained on the entire duration of non-failure and failure periods. We assessed their performance using only the first \textit{n} seconds of the failure period. Figure \ref{fig2} illustrates the average accuracy of the models in distinguishing between non-failure (NF) and executional failure (EF), as well as between non-failure and decisional failure (DF). As \textit{n} increases to 5 seconds, the accuracy stabilizes. For distinguishing EF from NF, the accuracy remains around 90\%, with a Recall of Failure of approximately 94\% across all classifiers. Similarly, for distinguishing DF from NF, the accuracy stabilises around 80\%, with a Recall of Failure of approximately 90\% for all classifiers.

\subsection{Evaluating Classification Performance for Real-Time Failure Detection}

\begin{figure*}[h!]
  \centering
  
  \begin{subfigure}[b]{0.3\linewidth}
    \centering
    \includegraphics[width=\linewidth]{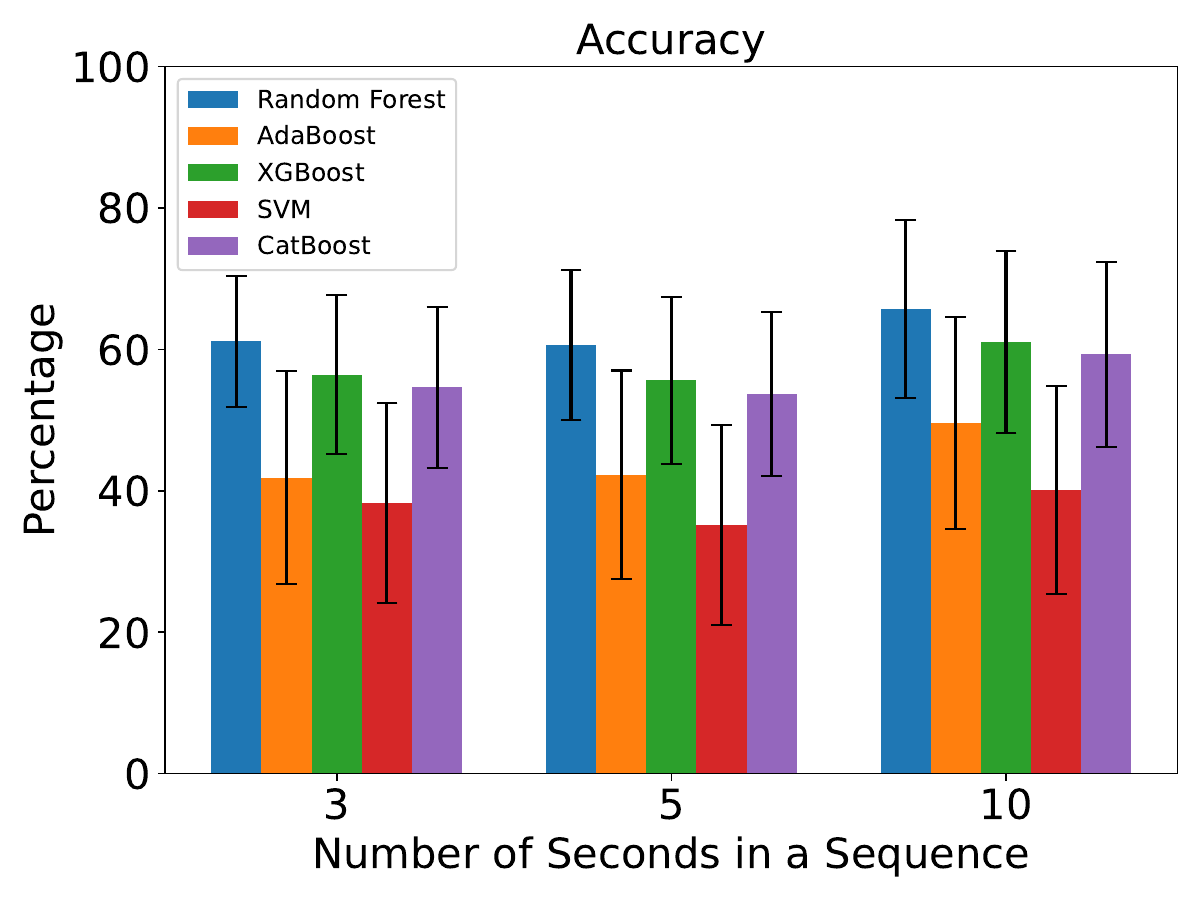}
    \caption{ }
    \label{fig3a}
  \end{subfigure}
  \begin{subfigure}[b]{0.3\linewidth}
    \centering
    \includegraphics[width=\linewidth]{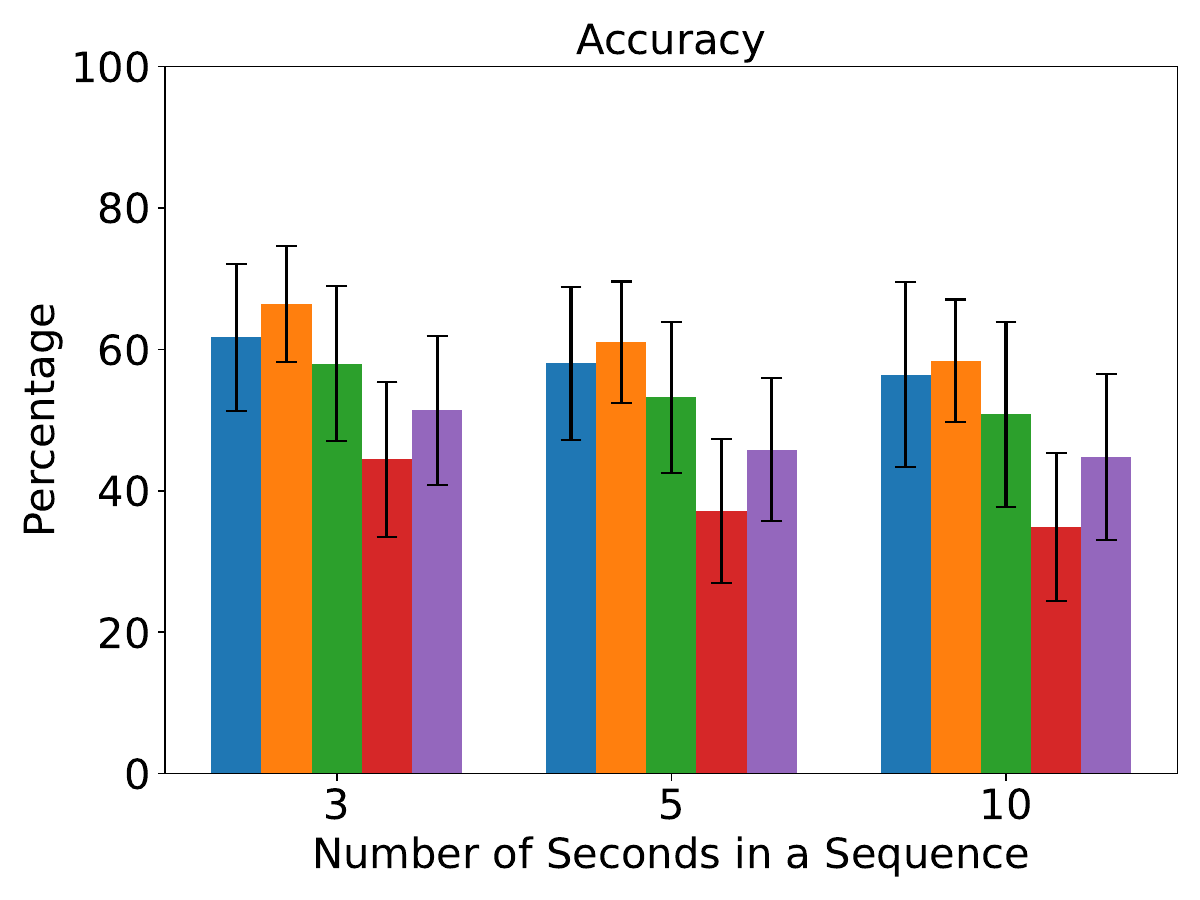}
    \caption{ }
    \label{fig3b}
  \end{subfigure}\\
  \begin{subfigure}[b]{0.3\linewidth}
    \centering
    \includegraphics[width=\linewidth]{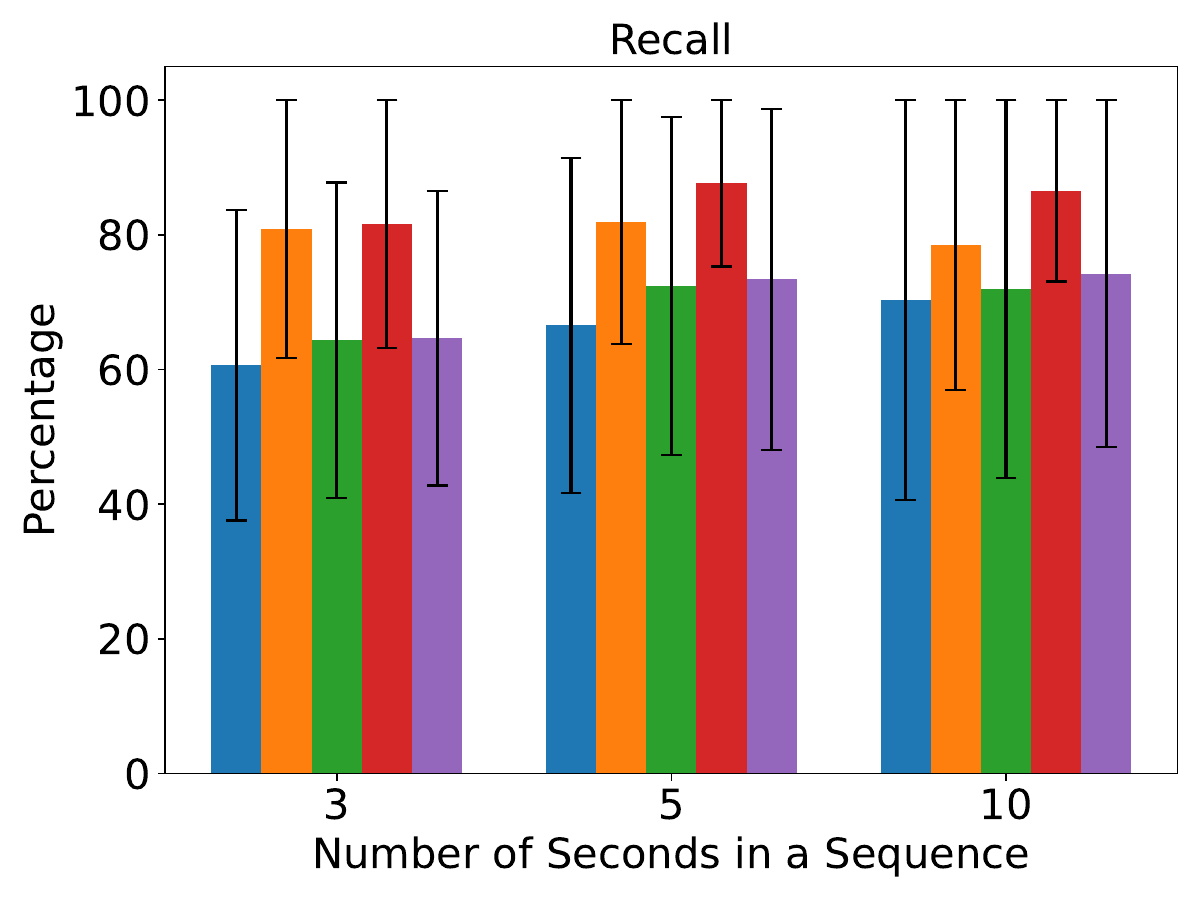}
    \caption{ }
    \label{fig3c}
  \end{subfigure}
  \begin{subfigure}[b]{0.3\linewidth}
    \centering
    \includegraphics[width=\linewidth]{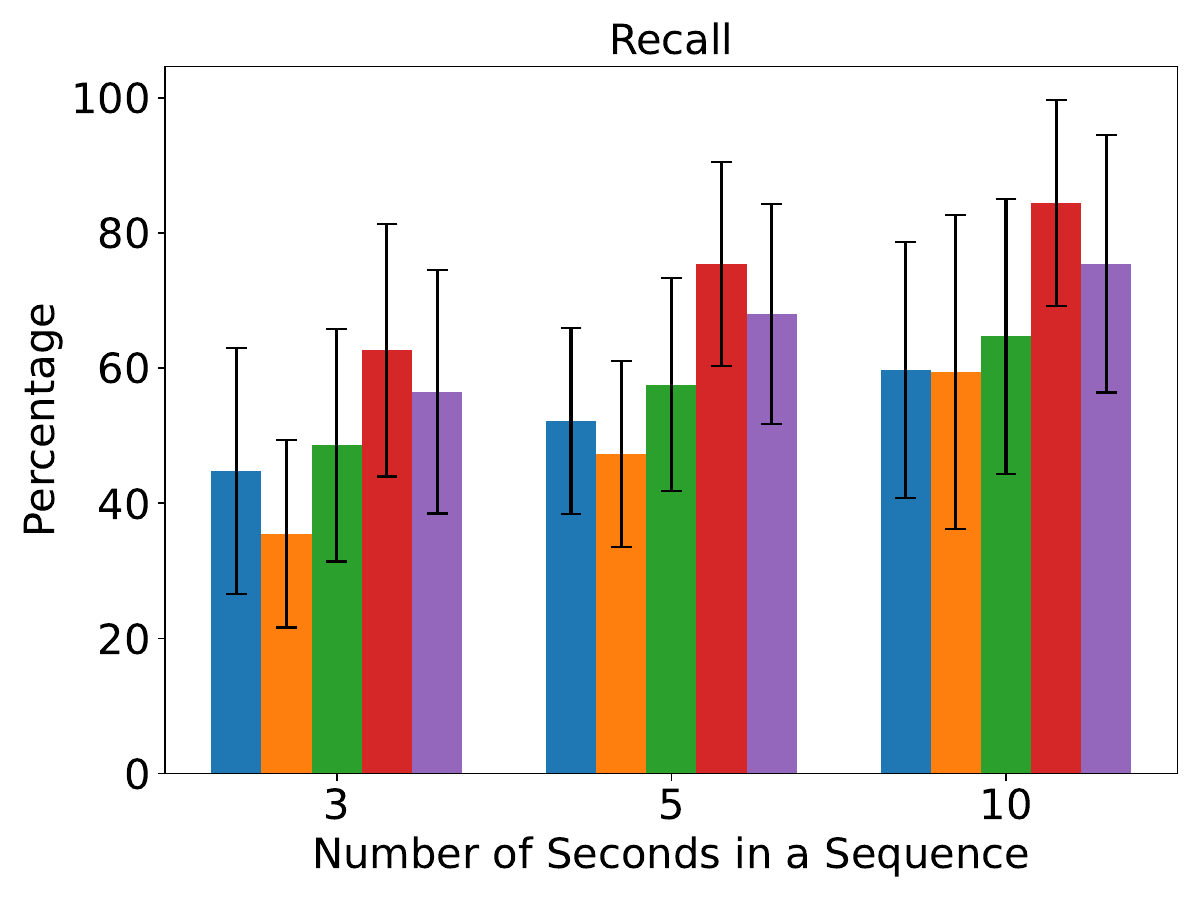}
    \caption{ }
    \label{fig3d}
  \end{subfigure}
    
  \caption{Classifier performance evaluated using eye-tracking metrics segmented into intervals of 3, 5, and 10 seconds with a 1-second sliding window: a) Accuracy in distinguishing between non-failure and executional failure, b) Accuracy in distinguishing between non-failure and decisional failure, c) Recall in detecting executional failure, and d) Recall in detecting decisional failure.}
  \label{fig3}
\end{figure*}

To enable the robot to detect its mistakes in real-life scenarios, it needs to repeatedly check at regular intervals whether something has gone wrong. In this section, we aim to address both research questions. In addition to the machine learning models used in the previous section, we also include SVM and CatBoost here.

The models were trained on the entire duration of non-failure and failure periods and evaluated by segmenting the eye-tracking metrics data into intervals of 3, 5, and 10 seconds, using a sliding window of 1 second. Figures \ref{fig3a}, \ref{fig3b}, \ref{fig3c}, and \ref{fig3d} illustrate the performance of the models in distinguishing between NF and EF and between NF and DF.

For both NF vs. EF and NF vs. DF, Random Forest achieved the highest accuracy compared to other classifiers, with an accuracy of approximately 60\%. In contrast, SVM had the lowest accuracy, below 50\%. However, SVM was the most effective in detecting the highest number of failures for both failure types.

  
  

\begin{figure*}[h!]
  \centering
  
  \begin{subfigure}[b]{0.3\linewidth}
    \centering
    \includegraphics[width=\linewidth]{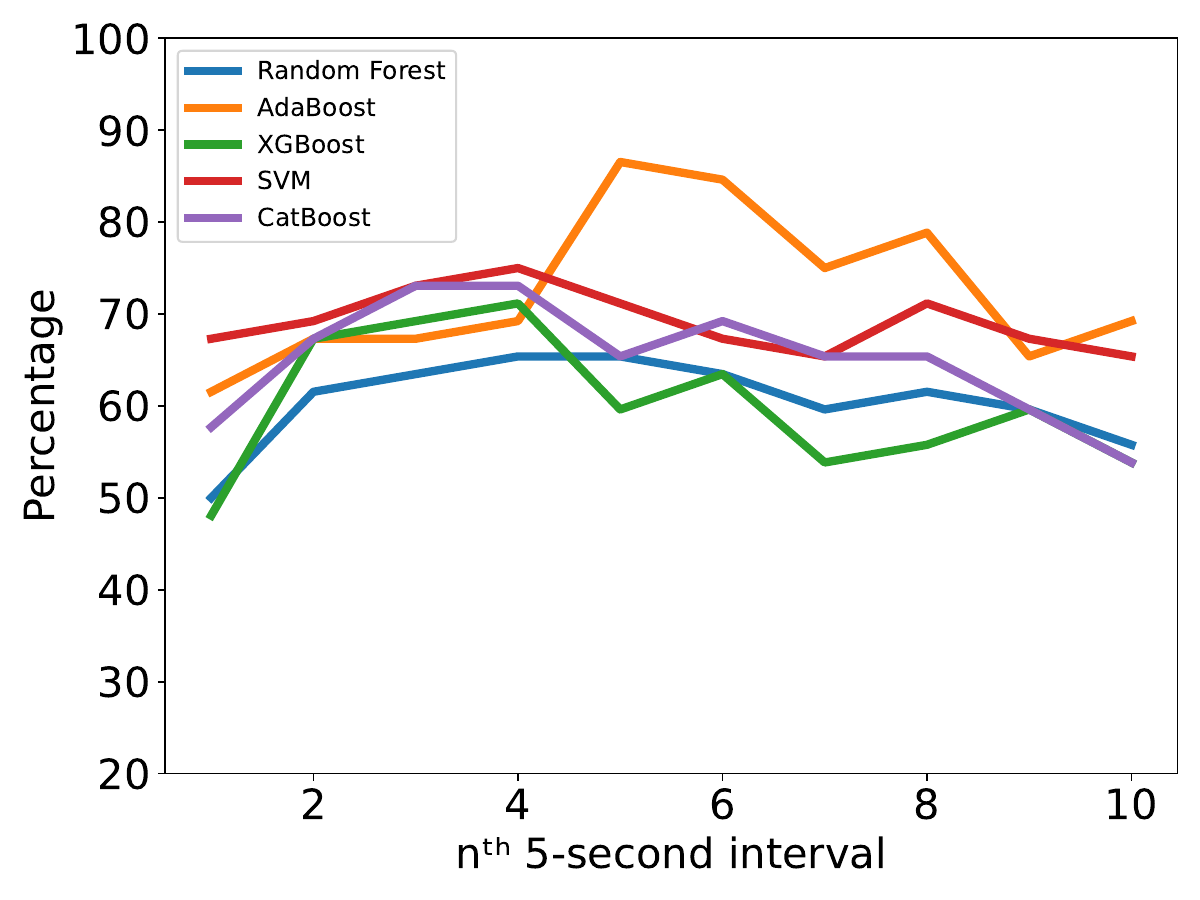}
    \caption{ }
    \label{fig3e}
  \end{subfigure}
  \begin{subfigure}[b]{0.3\linewidth}
    \centering
    \includegraphics[width=\linewidth]{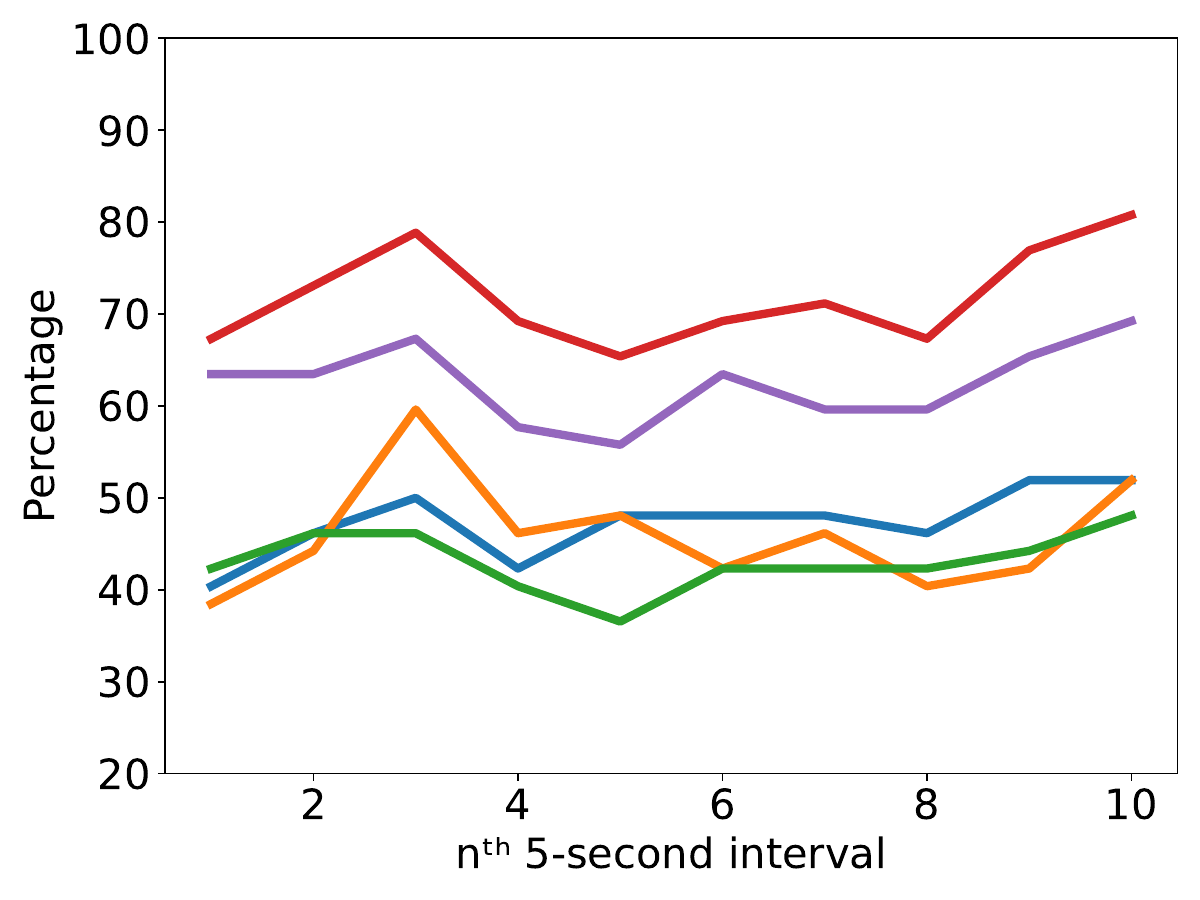}
    \caption{ }
    \label{fig3f}
  \end{subfigure}
    
  \caption{Percentage of users for whom the model successfully detected failures during each 5-second interval of the failure phase: a) Executional Failure, and b) Decisional Failure.}
  \label{fig3}
\end{figure*}



Additionally, we calculated the percentage of users, during each 3-, 5-, and 10-second interval of the failure phase, for whom the model successfully detected the failure. Figure \ref{fig3e} shows these percentages for EF at the 5-second interval, and Figure \ref{fig3f} shows them for DF at the same interval.

The results showed that, for EF, the models were most accurate at detecting user reactions during the period from 4 to 7 seconds for the 3-second interval, and from 3 to 8 seconds for the 5-second interval after a failure began. Similarly, for DF, the optimal detection period occurred from 2 to 5 seconds for the 3-second interval, and from 1 to 6 seconds for the 5-second interval after a failure began.

%% file: sections/05_Discussion.tex
\section{DISCUSSION and CONCLUSION}  
This study highlights the importance of user gaze dynamics in detecting robot failures during collaborative tasks. The results demonstrate that gaze-based machine learning classifiers can identify robot errors with high accuracy when the models are trained by labelling each pick-and-place action as either a failure or a non-failure, and tested similarly while reducing the duration of failure periods. However, since the exact moment of a robot failure is unknown, the robot needs to repeatedly analyse user gaze at regular intervals to determine whether a failure has occurred. For this purpose, we tested time intervals of 3, 5, and 10 seconds.
Although this real-time approach does not achieve the same level of accuracy as the previous method, it allows robots to continuously monitor for EFs or DFs. The models were more effective at distinguishing between NF and EF than between NF and DF. Among the classifiers, Random Forest achieved the highest accuracy, but its recall of failures was lower than others. For higher recall rates, SVM may be a better option as it detects the highest number of failures. 

Additionally, we analysed the percentage of users whose failures were correctly detected within each 5-second interval after a failure began. For EF, this percentage was approximately 70\%, while for DF, the results varied across classifiers, with CatBoost reaching around 60\%. Unlike user facial expressions in response to failures, as studied in \cite{stiber_using_2023}, gaze reactions do not exhibit specific characteristics like reaction time and reaction duration. As shown in the results, varying the duration of failure periods or using different 5-second intervals yielded consistent performance across models.

Despite these promising results, several limitations remain. In some cases, the robot's actions, such as placing its piece, overlapped with participants planning their next move, which could affect model accuracy. Furthermore, the study relied solely on gaze behaviour as an error indicator. Integrating multimodal cues, such as facial expressions, body movements, and speech, could enhance detection accuracy and robustness.

In conclusion, leveraging user gaze dynamics for robot error detection represents a significant step toward improving the reliability and trustworthiness of collaborative robots. This approach has the potential to enhance human-robot collaboration by enabling robots to proactively detect and recover from errors in real-time.